# OPTICAL VORTEX TORQUE MEASURED WITH OPTICALLY TRAPPED MICROBARBELLS

**Weronika Lamperska[1,*], Jan Masajada[1], Sławomir Drobczyński[1], Piotr Wasylczyk[2]**

[1]*Department of Optics and Photonics, Faculty of Fundamental Problems of Technology, Wroclaw University of Science and Technology, Wybrzeże Wyspiańskiego 27, PL50370 Wrocław, Poland*
[2]*Photonic Nanostructure Facility (PNaF), Faculty of Physics, University of Warsaw, ul. Pasteura 5, PL02093 Warszawa, Poland*
*Corresponding author: weronika.lamperska@pwr.edu.pl

**Optical vortex beams carry orbital angular momentum and thus exert torque on illuminated objects. A dielectric microtool – a microbarbell – is used in a two-laser optical tweezers to measure the torque of a focused optical vortex. The tool was either freely rotating due to the applied torque or set into oscillations by the counteracting force. Four different trapping configurations provided different ways of sensing the torque and gave consistent results. The value of torque was determined by confronting the experimental results with numerical and analytical models.**

Light can exert torque on matter due to angular momentum transfer [1,2] - either spin angular momentum (SAM) originating from elliptical and circular polarization state, or orbital angular momentum (OAM) arising from the twisted wavefront. Torque measurements of both spin and orbital angular momentum components have been of interest for driving micro- and nanomotors [2,3] and in material science for characterization and manufacturing [4,5]. The first experiments focused on the demonstration of the OAM transfer [6-8] and later quantitative measurements were reported [9]. Optical vortex (i.e. a light beam carrying OAM) can be generated in holographic optical tweezers (HOT) and used to exert torque on microobjects. A few methods for the torque measurements in HOT or similar system have been proposed [3,10,11]. In some of them [10,11] microtools fabricated by two-photon polymerization were used. In these papers torque arising from the transfer of total angular momentum or linear momentum was measured in close proximity to the beam center. We designed, manufactured and tested a new microtool – a microbarbell – for torque measurements in optical tweezers. The microbarbell is driven by the high-order vortex beam, thus the torque resulting exclusively from the OAM is measured relatively far from the beam axis. Four different methods of the tool trapping were tested with a single microtool and compared in situ. Since linearly polarized beam is used, the influence of the spin component of the angular momentum is eliminated. The measurements of torque due to spin component were reported elsewhere [3,10,12].

A microbarbell (Fig. 1) consists of three identical spherical beads mounted on a bar. The dimensions of the components are shown in Fig. 1c. Microbarbells were fabricated with a 3-D two-photon laser lithography [13] (Photonics Professional, Nanoscribe [14]) in a UV-curable photoresist (IPL, Nanoscribe) on a glass coverslip spin-coated with a thin (<1 μm) layer of polyvinyl alcohol (PVA). A typical sample was comprised of a matrix of 6 to 12 microbarbells (Figure 1b). Shortly before use, a one-millimeter-high chamber was built on a glass substrate by covering the glass with another coverslip and filled with purified water. As the PVA layer dissolves in water, the microtools are released from the substrate.

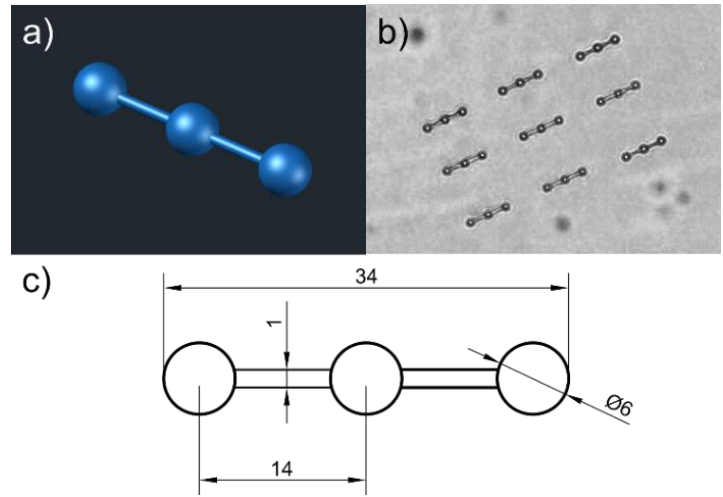

Fig. 1. (a) 3-D visualization of a microbarbell. (b) Optical microscope image of a matrix of 9 microbarbells laser-printed on a coverslip. (c) Dimensions of a microbarbell, in micrometers. The arm of the tool is the distance between the centers of side and central beads (14 μm).

Our microtools are operated in a two-laser optical tweezers. We have previously shown that the two-laser optical tweezers can be used for viscosity and trap stiffness measurements [15]. There are two optical paths to control optical traps in two different ways. The holographic path (HP) consists of Nd:YAG laser (Quantum Ventus, 1064nm, 4W) and reflective spatial light modulator (SLM, HoloEye-Pluto). The non-holographic path (NHP) is built of Nd:YAG laser (Quantum Ventus, 1064nm, 3W) and a system of 2 scanning galvanometer mirrors controlling the position of optical traps. In both paths half-wave plates are used to set two orthogonal linear polarization states of the beams. The beams are combined by a polarizing beamsplitter (PBS) and then focused by a high numerical aperture microscope objective (100x, oil-immersion, NA=1.3, Olympus UPLAN FL N). The sample is mounted on a heating stage maintaining the constant temperature of 25°C. Both lasers work in a constant work mode but they can be also connected independently to the waveform generator producing a square-wave signal of a 50% duty cycle at a set frequency, thus generating a blinking beam.

Four different experiments were performed and for each experiment the trap configuration was enriched with a new feature – a graphical overview of these configurations is shown in Fig. 3. The initial setup (Fig. 3a) consisted of two optical traps – a zeroth-order diffraction (ZOD) beam, so-called 'ghost trap' and a Laguerre-Gaussian trap, i.e. a vortex trap. Both traps were generated by the SLM so this configuration involves the holographic path only. The central bead of the microbarbell was trapped in the ZOD beam, while the side beads were illuminated by the optical vortex. Topological charge of the vortex $m$ indicates how fast the phase circulates around the beam axis and influences the diameter of the bright ring of the vortex beam. The topological charge of the vortex $m = 75$ was chosen to match its radius to the tool arm length (14 μm, Fig. 1c).

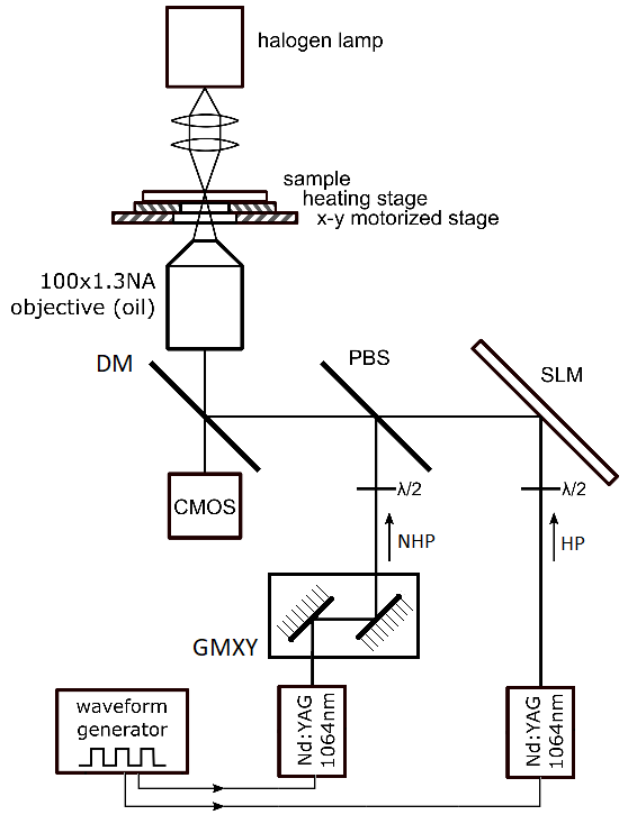

Fig. 2. Scheme of two-laser optical tweezers with a blinking beam. The holographic (HP) and non-holographic (NHP) paths are combined in the polarizing beamsplitter (PBS) and focused by the objective onto the sample. DM – dichroic mirror, GMXY - scanning galvanometer mirrors.

Since both traps were constant (non-blinking), the microbarbell experienced a constant torque from the optical vortex, resulting in a continuous circular motion (Visualization 1). The angular velocity depended on the laser power and the viscosity of the surrounding medium. In all experiments the vortex-generating laser power was set to 280 mW and the viscosity of medium (water, 25°C) was 0.890 mPa·s [16]. In these conditions the microbarbell rotated with angular velocity $\omega$ = (0.31-0.35) rad/s. To verify whether the rotary motion originated from the transfer of the orbital angular momentum, the handedness of the vortex was reversed by changing the topological charge from 75 to -75. The tool instantly started to rotate in the opposite direction, confirming the orbital angular momentum transfer.

In the second experiment (Fig. 3-II) both ZOD and vortex traps were blinking with the frequency of 1 Hz. The microbarbell rotated and stopped alternately, reminding a second hand of a mechanical clock. Both stages lasted for 0.5 second (half-period). The measured angular velocity was $\omega$ = (0.19-0.21) rad/s. Compared to the previous case, the velocity was not twice as low as one could expect. The reason was a slight movement of the microbarbell in z-axis caused by blinking of both traps. As the traps went off, the tool sank and then got lifted when the traps reappeared (Visualization 2).

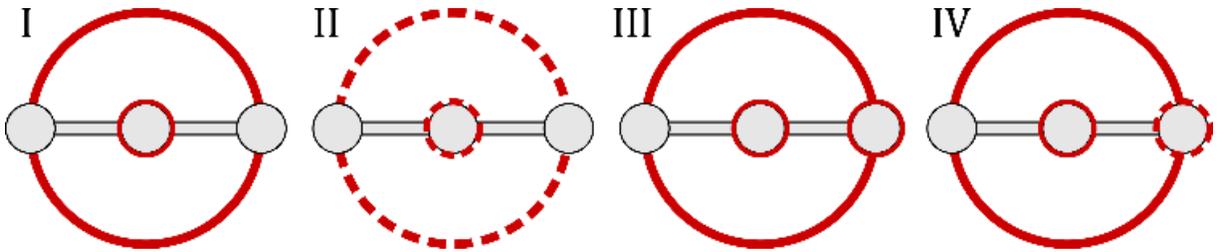

Fig. 3. A single microbarbell trapped in four different trap configurations (I-IV). Solid and dashed lines denote constant and blinking traps, respectively. The Gaussian beam trapping the right bead in III and IV is delivered by the non-holographic path (NHP). All the remaining traps are generated in the holographic path (HP).

In the third experiment (Fig. 3-III) a constant Gaussian trap from the non-holographic path was added to the original setup, illuminating one of the side beads in the microbarbell and serving as a counter-force for the optical vortex. Gaussian and vortex traps come from different laser sources which prevents their interference. The microbarbell was fixed in the stable position as long as the power of the non-holographic Gaussian trap exceeded the power of vortex trap. The power of NHP laser was lowered gradually until the point when the side bead escapes from the trap. The 'escape power' recalculated into trapping force was a measure of the vortex torque.

In the last experiment (Fig. 3-IV) a non-holographic Gaussian trap was blinking with a set frequency. The microbarbell was oscillating back and forth (Fig. 4a, Visualization 3) becoming a driven harmonic oscillator with a constant driving force (from the vortex beam) and a periodic restoring force (from the blinking Gaussian trap). The oscillations of the microbarbell were regular and followed the blinking frequency of the trap (Fig. 4b). The peak-to-peak amplitude depended mainly on the power of the vortex beam, whereas the shape of oscillations depended on the relative power of the vortex and the Gaussian (restoring) beam.

The data from the experiments: the angular velocity in configurations I and II, the 'escape power' in III and the amplitude of oscillations in IV, were used as input parameters for analytical and numerical models and allowed us to determine the torque of the optical vortex.

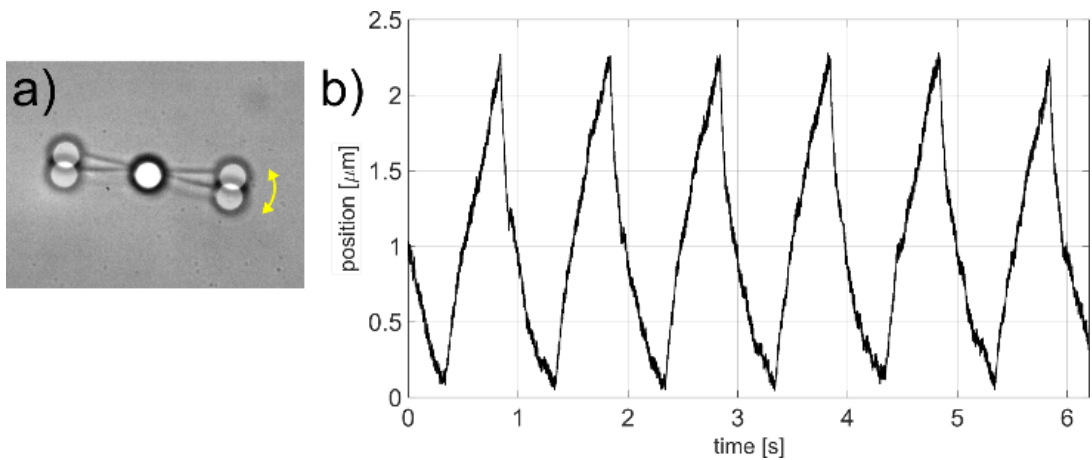

Fig. 4. a) Two superimposed photos of oscillating microbarbell and b) an example of the displacement of the side bead versus time.

The equation of motion for an oscillating side bead in the most general form is:

$$a\ddot{x} + b\dot{x} + cx + d = 0. \qquad (1)$$

The ratio of inertial term ($a\ddot{x}$) and viscous term ($b\dot{x}$) is the Reynolds number (Rn). Very low Reynolds number regime (Rn<<1) holds when viscous forces strongly outbalance the inertial ones. In optical tweezers one usually deals with microscopic object of small mass (~$10^{-12}$ - $10^{-16}$ kg) moving with low speed within relatively highly-viscous fluid. Therefore, due to very low Reynolds number approximation, the inertial term can be neglected [17-19]. The simplified Eq. 1 with the parameters (renamed for clarity) reads:

$$\alpha \dot{x} + \beta x + \gamma = 0. \qquad (2)$$

where $\alpha = -\dfrac{12\pi\mu r R}{I}$, $\beta = -\dfrac{kR}{I} g(t)$, $\gamma = \dfrac{F_{OV} R}{I}$, $\mu$ - viscosity of medium, $r$ – bead radius, $R$ – arm length of the microtool, $I$ – moment of inertia of the microtool, $k$ – trap stiffness, $F_{OV}$ – optical vortex force, $g(t)$ – 'blinking' function of period T (assuming square wave with 50% fill factor). The coefficient $\alpha$ accounts for the drag acting only upon side beads and not on the bar. It can be fixed by scaling the viscosity of the medium by a factor $c$ equal to the relative increase of the cross-sectional area multiplied by linear velocity after including the bar. For our microbarbell $c = 1.14$. and we replace $\mu$ with $\mu_c = c \cdot \mu$.

The solution of Eq. 2. can be written in a piecewise form:

$$x(t) = \begin{cases} -\dfrac{\gamma}{\alpha}t + x_0 & \text{for } 0 \leq t \leq \dfrac{T}{2} \\ -\dfrac{\gamma}{\beta} + \gamma\left(\dfrac{1}{\beta} - \dfrac{T}{2\alpha}\right)e^{\beta\left(\dfrac{T}{2\alpha} - \dfrac{t}{\gamma}\right)} & \text{for } \dfrac{T}{2} \leq t \leq T \end{cases} \qquad (3)$$

where $x_0$ is the final bead position at the end of each period calculated from the second equation. For the first period $x_0 = 0$.

The analytical model treats the bead as a single point and assumes infinite range of the restoring force. More realistic picture was achieved by solving the modified Eq. 2 numerically. The modification introduced Gaussian distribution of $k$ to the harmonic term and included the beam-bead overlap. The standard deviation σ of the Gaussian distribution was estimated from the trap geometry for a given laser power. Fig. 5 shows the experimental data and corresponding analytical and numerical solutions. Both solutions used identical parameters matched to the experimental conditions (e.g. temperature, viscosity of water, trap stiffness). The only parameter fitted was the force exerted by the optical vortex. The unrealistic assumptions of analytical model corrected further with the numerical one do not seem to affect the solution. However, the differences become apparent for very weak restoring force.

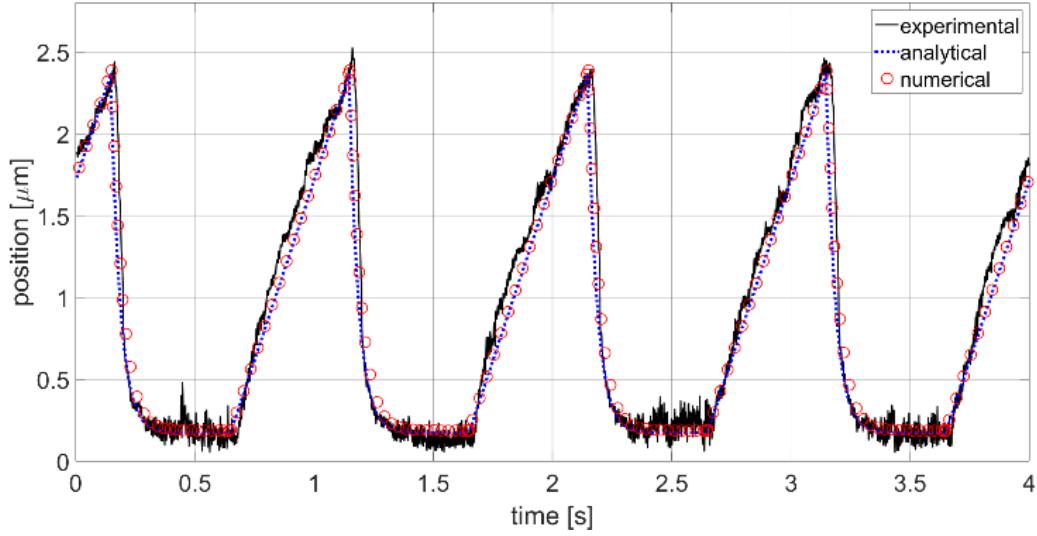

Fig. 5. Comparison of the experimental data (solid black line) with analytical (Eq. 3, dotted blue line) and numerical (red circles) results. Input parameters: $\mu_c = 1.18 \cdot 8.9\times10^{-04}$ Pa·s, r = $3\times10^{-6}$ m, R = $14\times10^{-6}$ m, $I = 4.9\times10^{-23}$ kg·m², k = $3.5\times10^{-6}$ N/m. Fitted parameter: $F_{OV}$=0.262 pN.

**Table 1. Mean values and standard deviations of the force ($F_{OV}$) and torque ($T_{OV} = R \cdot F_{OV}$) of the optical vortex measured in different traps configurations.**

|  | trapping configuration | | | |
| --- | --- | --- | --- | --- |
|  | I | II | III | IV |
| $F_{OV}$ [pN] (relative error) | 0.277 ± 0.021 (7.6%) | 0.325 ± 0.022 (6.8%) | 0.28 ± 0.17 (61%) | 0.255 ± 0.015 (5.9%) |
| $T_{OV}$ [×10⁻¹⁸ N·m] = [pN·µm] | 3,878 ± 0,294 | 4,550 ± 0,308 | 3,92 ± 2,38 | 3,570 ± 0,210 |

Table 1. summarizes the results of force and torque measurements. The experimental configurations can be divided into two groups: I and II are simple and direct, they use only one (holographic) optical path, rely on the direct observation of the angular velocity of the microtool and give the smallest relative errors. Yet, they suffer from strong susceptibility to any imperfections of the vortex geometry. A phase hologram generating an optical vortex of topological charge m = 75 is a structure with fine details (Fig. 6) and suffers from losses when displayed on a 8-bit 1920x1080 pixel SLM matrix. As a result, the intensity distribution of the vortex trap is non-

uniform and a rotating microbarbell experiences non-uniform light intensity across the vortex bright ring circumference which affects its motion. Methods III and IV are more resistant to imperfect vortex geometry because the oscillations of a tool occur only in a small section of the bright ring. On the other hand, the results obtained with method III are strongly dependent on the choice of σ (standard deviation of the Gaussian distribution in the non-holographic trap), hence the relative error of over 60%. In method IV the uncertainty of σ does not affect the outcome until the amplitude of the oscillations reaches its maximum, above which the bead is no longer restored by the Gaussian trap.

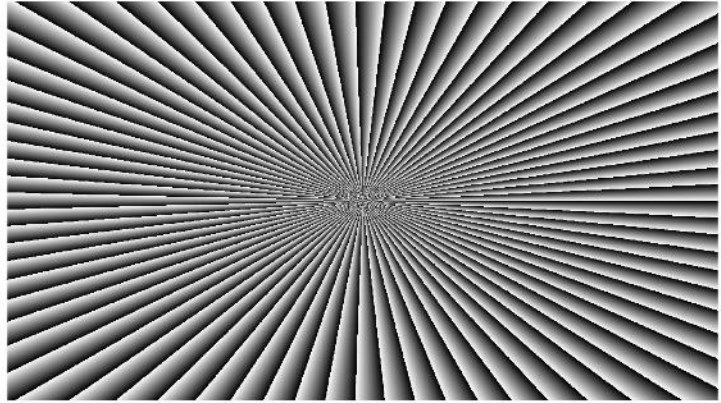

Fig. 6. Exemplary 8-bit grayscale fork hologram generating optical vortex of topological charge m = 75.

In this Letter we demonstrated a microbarbell - a 3D-printed microtool designed for optical metrology with optical tweezers. The microbarbell was used for measuring the torque exerted by a focused optical vortex. The dynamics of the system was formulated within the framework of a driven harmonic oscillator and solved both analytically and numerically. The solutions are in good agreement with experiment. Four different experimental configurations were used and compared in terms of difficulty, reliability of results and potential error sources. All the methods yielded comparable results for the optical vortex torque equal to (3.5÷4.5) pN·µm, consistent with the values obtained in other experiments [10,20]. The method based on inducing the microbarbell oscillations proved to be the least prone to systematic errors and the most precise. Both single-laser methods (I and II) are also promising yet require improvements, which can be achieved in two ways. Since the problem lies in poor quality of the vortex hologram, one could enhance its the resolution by replacing the Full HD SLM with a new-generation 4K UHD. Another solution is to shorten the microbarbell so that the optical vortex of smaller topological charge can be used. However, there is a limit below which the ZOD and the vortex traps start to interfere and disrupt the motion of the microbarbell. Method III is the most inaccurate as long as the exact width of the Gaussian beam is unknown.

**Funding.** Polish Ministry of Science and Higher Education (DI2016006146), European Regional Development Fund Operational Programme (POIG.02.01.00-14-122/09).

**Disclosures**. The authors declare no conflicts of interest.